\documentclass[pra,twocolumn,groupedaddress,amsmath,floatfix]{revtex4}
\usepackage{graphicx}% Include figure files
\usepackage{bm}

\newcommand{\ch}{\hat{c}}
\newcommand{\Tt}{T_{\rm t}}
\newcommand{\Ht}{H_{\rm t}}

\newcommand{\as}{a_s}

\newcommand{\vf}{v_{\rm F}}

\newcommand{\curH}{{\cal H}}
\newcommand{\curF}{{\cal F}}

\newcommand{\phdag}{{\phantom{\dagger}}}

\newcommand{\kf}{k_{\rm F}}

\newcommand{\curO}{{\cal O}}

\newcommand{\real}{{\rm Re}}

\newcommand{\bk}{{\bf k}}
\newcommand{\bq}{{\bf q}}
\newcommand{\bp}{{\bf p}}

\newcommand{\grad}{{\bm{\nabla}}}

\newcommand{\bx}{{\bf x}}

\newcommand{\be}{\begin{equation}}
\newcommand{\ee}{\end{equation}}
\newcommand{\bea}{\begin{eqnarray}}
\newcommand{\eea}{\end{eqnarray}}
\newcommand{\bse}{\begin{subequations}}
\newcommand{\ese}{\end{subequations}}

\input{epsf}
\begin{document}
%\draft
\title{Polarized superfluids near their tricritical point}
%-------------------------
\author{Daniel E.~Sheehy}
\affiliation{Department of Physics and Astronomy, Louisiana State University, Baton Rouge, Louisiana 70803}
\date{July 6, 2008; Revised Dec. 30, 2008}
%\maketitle
%--------------------------
\begin{abstract}
%--------------------------
We obtain a variety of predictions for the properties of population-imbalanced (or polarized) 
fermionic superfluids near their tricritical point. 
In the vicinity of the high-symmetry tricritical point, observable quantities  such as the cloud shape, heat capacity, local
polarization and correlation length should exhibit distinct behavior arising from the tricritical scaling laws, as well as logarithmic corrections to 
scaling reflecting the marginal nature of interactions.  
%--------------------------
\end{abstract}
\maketitle
%--------------------------

\section{Introduction}
\label{SEC:intro}
One of the most exciting recent developments in atomic physics has
been the achievement of paired superfluidity of fermionic atomic 
gases~\cite{regal2004,zwierlein2004,kinast2004,bourdel2004,chin2004,Partridge05p,zwierlein2005}.
Such superfluidity, arising from attractive interactions between two fermion species
mediated by a magnetic-field tuned Feshbach resonance, can be continously experimentally 
tuned (by adjusting the external magnetic field)  from the BCS limit of weak pairing to 
the BEC limit of strong pairing~\cite{GR07,Giorgini08}. 

In fact, the smooth crossover between these two limits only occurs for an {\it equal number\/} of
the two fermion species and any population imbalance (or polarization) interrupts the
BEC-BCS crossover, as seen in recent experiments~\cite{MIT,Rice1,Rice2,Shin07}. 
Theoretical work~\cite{Liu,Bedaque2003,SR2006,Gubbels,Parish2007,Chien,pilati} on such polarized 
Fermi gases predicts a rich ground-state phase diagram including
an FFLO phase~\cite{FFLO} in the positive detuning BCS regime,
a strongly-interacting polarized normal Fermi gas at large polarization~\cite{lobo2006}, a regime of magnetic 
superfluidity (consisting of tightly-bound molecules coexisting with a single-spin Fermi sea, analogous
to $^3$He-$^4$He mixtures~\cite{BEG,bulgac2007}), and a regime of phase separation.

\begin{figure}%
\epsfxsize=9cm
\vskip.3cm
\centerline{\epsfbox{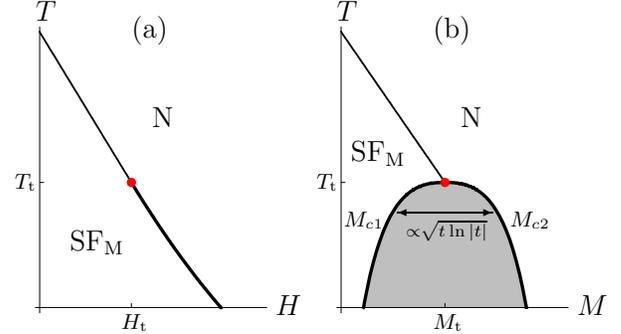}}
\vskip-.75cm
\caption{
(Color Online) 
(a) Phase diagram vs. chemical potential
difference $H$ and temperature $T$ near the tricritical point (TP) at $H_{\rm t},T_{\rm t}$ (solid point), showing 
magnetic superfluid (SF$_{\rm M}$) and polarized Fermi liquid (N) regions separated by a 
continuous transition at $T>T_{\rm t}$ (thin line) and a first-order transition at $T<T_{\rm t}$ (thick line). 
 (b) Phase diagram vs. magnetization $M$ and $T$, near the TP at $M_{\rm t},T_{\rm t}$, 
showing the coexistence region (shaded; bounded by $M_{c1}$ and $M_{c2}$), with the coexistence curve asymptotically satisfying 
$M_{c2} -  M_{c1} \propto \sqrt{t\ln|t|}$ with $t \equiv (T-\Tt)/\Tt$.}
\label{phasediagram}
\vskip-.35cm
\end{figure}

Although a consistent picture of the ground-state phase diagram has arisen from mean-field theory~\cite{SR2006,Parish2007}
and Monte Carlo~\cite{pilati} results, the finite-temperature phase diagram~\cite{Gubbels,Parish2007,Chien} is less well understood,
 particularly 
near the unitary point where the interspecies scattering length $\as$ diverges.  For any detuning, 
at sufficently low polarization or chemical potential difference $H$  we expect a finite-temperature 
phase transition at which the superfluid order parameter vanishes {\it continuously}.  Conversely, at low temperature,
superfluidity is destroyed in a first-order fashion~\cite{Clogston} with increasing $H$.  Across this phase transition at fixed $H$, the 
polarization or magnetization $M = n_\uparrow - n_\downarrow$  (with $n_\sigma$ the density of spin-$\sigma$ fermions) jumps discontinously;
at fixed $M$ one finds  a regime of phase separation~\cite{Bedaque2003,SR2006}, as seen in Fig.~\ref{phasediagram}b.
It is natural to suppose that these first and second-order regimes are connected by a tricritical point~\cite{Griffiths70} (TP), as 
which occurs in BCS superconductors under an imposed magnetic field and in $^3$He-$^4$He mixtures.

%------------------------------
%------------------------------

Indeed, recent experiments~\cite{Rice2,Shin07} have reported evidence for a TP in the unitary regime, so that
an exploration of additional tricritical phenomenology is
 of considerable interest.  Narrow-resonance models~\cite{Parish2007} predict that a line of tricritical points
crosses the phase diagram as a function of Feshbach resonance detuning, terminating at a quantum tricritical point
in the deep BEC limit~\cite{Parish2007}.   However, in the experimentally-relevant
wide-resonance limit, it is possible that strong correlations may interrupt the tricritical point,
as found in the Blume-Emery-Griffiths model~\cite{BEG}.  Thus, our aim is to devise novel experimenal signatures
of a tricritical point, to aid in establishing the phase diagram of polarized Fermi gases. 

  We thus proceed by assuming that the phase diagram of polarized superfluids 
possesses a TP.  If so, then, quite generally, the phase diagram near such a tricritical point
will resemble Fig.~\ref{phasediagram}, neglecting the possibility of an FFLO state, which, at least within 
mean-field theory, is restricted to a thin window of $H$ or $M$  values~\cite{SR2006,Parish2007} in the BCS regime.
Having made this assumption, we shall make predictions for the behavior, near the TP, of various observable quantities in
cold atom experiments, such as the local polarization (magnetization), molecular or pair density, and the heat capacity.
  Our results are based on 
the analysis of a  sixth-order Ginzburg-Landau (GL) free energy via 
mean-field theory and the renormalization group (RG) and are expected to apply generally to finite-temperature 
TPs in polarized Fermi gases~\cite{quantumtrinote}. 

This manuscript is organized as follows.  In Sec.~\ref{SEC:model}, we introduce the GL model of tricritical points and
discuss our principal results.  In Sec.~\ref{SEC:microscopic}, we derive the coefficients of the GL model within
a mean-field analysis of the one-channel model of resonantly-interacting polarized Fermi gases.  In Sec.~\ref{SEC:RG}, we derive the 
RG equations for the sixth-order GL model.  In Sec.~\ref{SEC:molecular} we use the RG to derive equations for the variation
of the superfluid order parameter below the phase transition in the vicinity of the tricritical point and show how these results imply a characteristic
steep cloud shape in a trapped polarized gas.  In Sec.~\ref{SEC:mag}, we derive a prediction for the jump in the magnetization $M$ across the first-order
phase boundary near the tricritical point, a quantity that translates, in a trapped polarized superfluid, to a jump in $M$ as a function of radius.  
In Sec.~\ref{SEC:heat} we derive equations for the heat capacity above and below the transition near the tricritical point.  In Sec.~\ref{SEC:corr},
we describe the divergence of the order parameter correlation length near the tricritical point before concluding in Sec.~\ref{SEC:concl}. 

\section{Model and principal results}
\label{SEC:model}
As we have noted,
a tricritical point in the phase diagram of polarized Fermi gases separates first-order 
and continuous phase transitions of the superfluid order parameter $\psi$, with 
$|\psi|^2$ essentially representing the density of condensed molecular pairs~\cite{densitynote}.
The behavior near the tricritical
point can be captured using the following sixth-order free-energy functional:
\be
F = \int d^3 x \Big[
\frac{\hbar^2|\grad \psi|^2}{2m_b} +\frac{1}{2}r|\psi|^2 + \frac{1}{4}u|\psi|^4 + 
\frac{1}{6}v |\psi|^6
\Big],
\label{eq:freeoriginal}
\ee
where $m_b = 2m$ is the molecular mass,  and $r$, $u$ and $v$ are $T$ (temperature) and $H$ (chemical potential difference) dependent
coefficients.  

 Below, we show how Eq.~(\ref{eq:freeoriginal}) can  be obtained within a mean-field analysis of the standard one-channel
model of resonantly-interacting Fermi gases, allowing us to derive mean-field predictions for the coefficients
$r$, $u$, $v$.  Although mean-field theory breaks down in the unitary regime, we nonetheless
expect Eq.~(\ref{eq:freeoriginal}) to correctly describe the vicinity of the tricritical point, but with unknown coefficients.
In the present section, we proceed by reviewing the mean-field phase diagram of Eq.~(\ref{eq:freeoriginal}).

%------------------------------
\begin{figure}%
\epsfxsize=9cm
\vskip.3cm
\centerline{\epsfbox{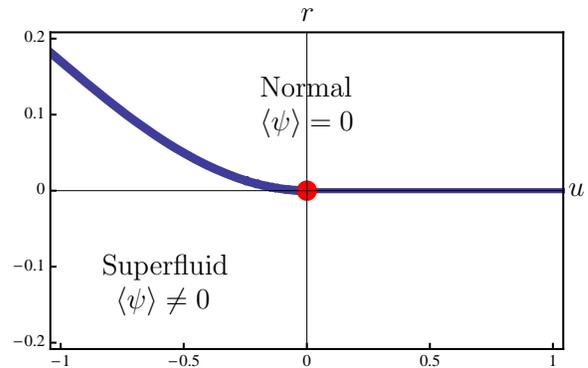}}
\vskip-.75cm
\caption{(Color Online)
Phase diagram of $F$ (with $v=1$), Eq.~(\ref{eq:freeoriginal}), showing superfluid and normal (nonsuperfluid)
phases separated by a first-order boundary for $u<0$ and a continuous boundary for $u>0$ with the 
tricritical point at the origin.}
\label{ApproxTriPhaseDia}
\vskip-.35cm
\end{figure}
%------------------------------

The basic tricritical phenomenology (including the phase diagram in the $u-r$ plane, Fig.~\ref{ApproxTriPhaseDia})
follows from analyzing Eq.~(\ref{eq:freeoriginal}) in the mean-field
approximation, by minimizing with respect to an assumed spatially-uniform $\psi$.
With a spatially uniform $\psi$, the mean-field free energy is:
\be
F = \frac{1}{2} r|\psi|^2 + \frac{1}{4} u |\psi|^4 +  \frac{1}{6} v |\psi|^6, 
\label{eq:meanfieldfree}
\ee
where we note that, in going from Eq.~(\ref{eq:freeoriginal}) to Eq.~(\ref{eq:meanfieldfree}), we
have clearly taken the system volume to be unity.  
Along with the trivial stationarity condition $\psi = 0$, one finds  the nontrivial stationarity condition
\be
 0 = r+ u|\psi|^2 + v|\psi|^4,
\label{eq:stationarity}
\ee
that can be solved to yield 
\be
|\psi|^2 = \frac{1}{2v} \big[ 
\sqrt{u^2-4rv} -u
\big].
\label{eq:themeanfieldresult}
\ee
We begin with the regime, $u>0$, in which the phase transition is continuous, that occurs for low $H$.
Since we expect $v>0$ (necessary to stabilize the tricritical point), we see that Eq.~(\ref{eq:themeanfieldresult}) 
only represents a physical solution for $r<0$, so that the stable solution is $\psi = 0$ for $r>0$.  This
nonsuperfluid, or normal, regime is characterized by a vanishing pair (or molecular) density 
$n_m = |\psi|^2$.  For $r<0$, in the $u>0$ regime superfluidity emerges continuously.
For small $r$ we see that $n_m \simeq -r/u$ is essentially proportional to the molecular chemical potential $\mu_m \equiv -r/2$, vanishing along $r=0$.

With increasing $H$, as the TP is approached, $u$ (proportional
to the molecular scattering length) approaches zero, so that the onset of superfluidity at the continuous transition becomes 
progressively steeper.
 The point $r=u=0$ defines the TP $(\Tt,\Ht)$.  
For $T<\Tt$ ($u<0$), the superfluid to normal transition is first order, with $n_m$ and the magnetization $M$ jumping 
discontinously at the phase boundary $r= 3u^2/16v$, yielding phase separation at fixed $T$ and $M$ below
$\Tt$ (Fig.~\ref{phasediagram}b).

Close to the TP, however, fluctuations of $\psi$ yield important modifications to the mean-field picture, best
captured via an RG analysis, in the form of logarithmic corrections to 
scaling~\cite{Larkin69,WR72,Stephen75,Lawrie} (already observed in a different setting~\cite{Shang80}).
For example, consider the molecular density 
along a line intersecting the TP (i.e., $H = H_t$ and $u=0$).  Along such a line,
mean-field theory predicts  $n_m = \sqrt{-r/v}$, which, along with $r \propto T-\Tt$, yields
$n_m \propto |t|^{2\beta}$ with $t\equiv \frac{T-\Tt}{\Tt}$ and the tricritical exponent $\beta = 1/4$, 
in contrast to $\beta = 1/2$ for the standard superfluid transition.  In fact, the predicted
onset is even steeper when fluctuations are accounted for, with
\be
n_m \propto \sqrt{t\ln|t|}.
\label{eq:nmintro}
\ee
Thus, fluctuations do not alter the mean-field exponent but instead provide a logarithmic factor;
this occurs because the tricritical upper critical dimension equals the physical dimension 
$d=3$~\cite{Larkin69,WR72,Stephen75,Lawrie,Shang80}.

%------------------------------
\begin{figure}%
\epsfxsize=10cm
\vskip.5cm
\centerline{\epsfbox{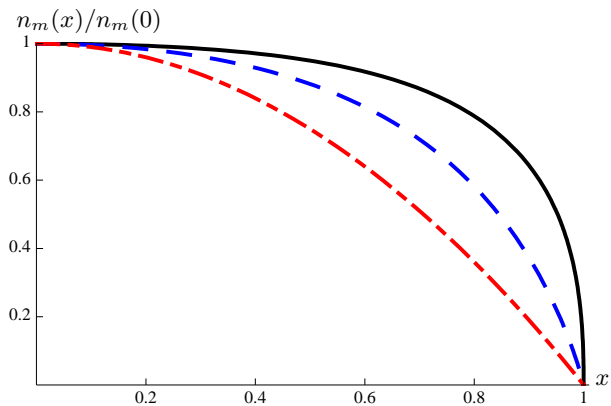}}
\vskip-.75cm
\caption{(Color online) Plots of the molecular density $n_m$ vs. position $x$ (normalized 
to reach unity at the edge), within the local density approximation [by replacing $r \to (1-x^2)$ in Eq.~(\ref{eq:nmresult})], for $v=1$  with
$u=0$ (tricritical case, solid curve) and $u=1$ (above the tricritical point, dashed curve).  For comparison, 
the  dot-dashed curve shows a standard parabolic Thomas-Fermi profile $n_m\propto (1-x^2)$.
  }\label{fig:densityplot}
\vskip-.35cm
\end{figure}
%------------------------------

In a cold atom experiment, the tricritical exponent $\beta = 1/4$  (and the logarithmic correction) should also be reflected
in the spatial dependence of the pair density in a harmonic trap which is captured, within the local density 
approximation (LDA), via the replacement $\mu_m \to \mu_m - 
\frac{1}{2} m_b \Omega^2 x^2$ (with $\Omega$ the trap frequency).  In a usual superfluid this yields the 
well-known parabolic Thomas-Fermi (TF)  profile $n_m \propto (1-x^2/x_{TF}^2)$ of a trapped gas (reflecting
the usual exponent $\beta = 1/2$), with $x_{TF}$ the TF radius.
In  the present  case of a polarized Fermi superfluid near the tricritical point, we find
 (setting $x_{TF}=1$ for simplicity):
\be
n_m(x) \simeq \sqrt{1-x^2} \ln^{1/2} (1-x^2)^{-1},
\label{Eq:cloudnearedge}
\ee
a distinctly different cloud shape (including a divergent slope 
at $x\to 1$) for polarized superfluid Fermi gases, as shown in Fig.~\ref{fig:densityplot}.  
The enhanced symmetry near the TP is also reflected in the heat capacity of 
this interacting Fermi gas~\cite{Kinast}; by considering the free energy 
for $t<0$ along with the logarithmic corrections, we find
\be
C\propto |t|^{-1/2}\ln^{1/2}|t|^{-1},
\label{eq:heatintro}
\ee
for the heat capacity below $\Tt$, again reflecting the mean-field tricritical heat capacity
exponent $\alpha = 1/2$ along with a logarithmic correction, in contrast to the behavior
of the standard superfluid transition~\cite{Barmatz}, described by a small exponent $\alpha \simeq -0.01$.
In the normal state, for $t>0$, we find $C\propto t^{-1/2}$,  
with no logarithmic correction.  We also find that the divergence of the correlation length near the phase transition 
(recently measured in a bosonic cold-atom system~\cite{Donner}), 
$\xi \propto |t|^{-\nu}$, reflects the correlation length exponent $\nu = 1/2$ with no logarithmic correction.

Tricritical scaling also implies a universal shape to the
coexistence  curve near $\Tt$, asymptotically equal to the jump in the magnetization 
$\delta M$ across the first-order phase boundary (recently studied near the TP by
the MIT group as a jump in the magnetization vs. radius~\cite{Shin07}); we find
(Fig.~\ref{phasediagram}b):
\be
\label{eq:magjump}
\delta M \propto \sqrt{t\ln |t|}.
\ee
The preceding expressions apply for uniform polarized Fermi gases close to the TP 
(and trapped Fermi gases, within the LDA, that have part of their system locally at the TP.)
Away from the TP, however,  such observables will  cross over 
to critical (for  $H<\Ht$) or first order  (for $H>\Ht$) behavior asymptotically close to the transition.  

\section{Microscopic Model}
\label{SEC:microscopic}
To derive these results, we begin with 
the standard one-channel model of two-component resonantly-interacting fermions
$\ch_{\bk\sigma}^\phdag$ (with $\sigma = \uparrow,\downarrow$): 
\be
\curH = \sum_{\bk,\sigma}\xi_{k\sigma}\ch_{\bk\sigma}^\dagger  \ch_{\bk\sigma}^\phdag 
+ \lambda\sum_{\bk\bq\bp} \ch_{\bk\uparrow}^\dagger \ch_{\bp\downarrow}^\dagger 
\ch_{\bk+\bq\downarrow}^\phdag
\ch_{\bp-\bq\uparrow}^\phdag,
\label{eq:singlechannelintro}
\ee
where $\xi_{k\sigma} = \epsilon_k - \mu_\sigma$, $\epsilon_k = k^2/2m$, and $m$ is the fermion mass.
 The population imbalance is induced by the difference in chemical potentials $\mu_\sigma = \mu \pm H$,
and  the attractive Feshbach resonance interaction is captured with the coupling 
constant $\lambda<0$, the magnitude of which increases with decreasing Feshbach resonance detuning,
and which is best characterized by its connection 
\be
\frac{m}{4\pi \as\hbar^2} 
= \frac{1}{\lambda}+\sum_\bk\frac{1}{2\epsilon_k},
\label{eq:scatt}
\ee
to the scattering length $\as$. 

We now proceed to focus on the TP 
of Eq.~(\ref{eq:singlechannelintro}) at finite temperature and polarization.  We do this by first mapping
Eq.~(\ref{eq:singlechannelintro}) onto the general tricritical free energy Eq.~(\ref{eq:freeoriginal}) thereby
deriving the coeffcients $r$, $u$, and $v$. This involves making the BCS mean-field approximation
 $\Delta(\bx) = \lambda \langle  \ch_{\downarrow}^\phdag(\bx) \ch_{\uparrow}^\phdag(\bx)\rangle$, 
expanding to leading order in the magnitude and spatial gradients of $\Delta(\bx)$, 
and tracing over the fermionic degrees of freedom in the partition function following standard methods~\cite{agd} 
(generalized to $H\neq 0$).  The inaccuracy of the BCS mean-field approximation near unitarity
 means our predictions
for the location of the tricritical point, 
and the precise  forms of $r$, $u$, and $v$ are not quantitatively 
trustworthy; however, we expect them to be qualitatively
valid (and, they must be if a tricritical point indeed occurs).  Moreover, our principal interest concerns
the power laws and logarithmic corrections near the tricritical point, that are independent of the BCS
approximation.   

With these caveats we begin by first assuming uniform $\Delta$; the gradient term in
Eq.~(\ref{eq:freeoriginal}) will be derived below.  With uniform $\Delta$, we can easily trace over
the fermion degrees of freedom to obtain the following mean-field free-energy:
\bea
&&F = - \frac{|\Delta|^2}{\lambda} - \sum_\bk (E_k - \xi_k)
\\
&& \qquad 
- T\sum_\sigma \sum_\bk \ln \big(
1+ {\rm e}^{-\beta(E_k + \sigma H)}
\big),
\nonumber
\eea
where $E_k = \sqrt{\xi_k^2 + |\Delta|^2}$.  Using Eq.~(\ref{eq:scatt}), Taylor expanding order by order in $\Delta$, 
and evaluating the momentum integrals yields:
\be
F = F_0 + V_2 |\Delta|^2 + \frac{1}{2} V_4 |\Delta|^4 + \frac{1}{3} V_6 |\Delta|^6,
\label{eq:fexpand}
\ee
where 
\bea
&&\hspace{-.5cm}V_2 \simeq \!\frac{m}{4\pi \as} +
\rho_0 \Big(\ln \frac{2\mu {\rm e}^{-2}}{\pi T} 
- \real \big[\psi\big(\frac{1}{2} + \frac{iH}{2\pi T}\big)\big]
\Big),\label{Eq:veetwo}
\\
&&\hspace{-.5cm}V_4 \simeq  -\frac{\rho_0}{16 \pi^2 T^2}\real\big[
\psi_2\big(\frac{1}{2} + \frac{iH}{2\pi T}\big)\big],  \label{Eq:veefour}
\\
&&\hspace{-.5cm}V_6 \simeq  \frac{\rho_0}{1024\pi^4 T^4} \real\big[
\psi_4\big(\frac{1}{2} + \frac{iH}{2\pi T}\big)\big],
\eea
with $\psi_n(x)$ the polygamma function and $\real$ denoting
the real part.  In evaluating these integrals, we have assumed $\mu>0$ and
expanded the fermion density of states to leading order at the Fermi surface (yielding the 
factors of $\rho_0  =  \frac{m^{3/2} \mu^{1/2}}{\sqrt{2}\pi^2\hbar^3}$), an approximation
that breaks down in the deep BEC regime, where $\mu$ becomes negative. 

The tricritical point occurs when $V_2 = V_4 =0$.  Examining Eq.~(\ref{Eq:veefour}), 
we see that mean-field theory predicts a universal ratio between the chemical difference $\Ht$ and temperature
$\Tt$ at the tricritical point, given by $\real\big[\psi_4\big(\frac{1}{2} + \frac{iH}{2\pi T}\big)\big]=0$,
or,
\be
\frac{\Ht}{\Tt}  \simeq 1.91,
\label{Eq:htri}
\ee
valid at any detuning (within the preceding mean-field assumptions).  Equation~(\ref{Eq:htri})
can be combined with Eq.~(\ref{Eq:veetwo}) to obtain a prediction for the ratio $\Tt/\mu$
as a function of the scattering length $\as$ that gives 
\be
\frac{\Tt}{\mu}  \simeq  0.35,
\label{Eq:ttri}
\ee
at the unitarity point $a_s^{-1}=0$.

Our next task is to obtain the gradient term in the free energy.  We do this directly following 
the textbook derivation of the GL free energy (as discussed in Ref.~\onlinecite{agd}),
that essentially allows spatial variations in the quadratic coefficient.  Here, 
the only difference is the nonzero chemical potential difference $H$.  We obtain an
$H$-dependent coeffcient of the gradient term that, taking $H$ to be given by its value at
the tricritical point Eq.~(\ref{Eq:htri}), yields
\bea
&&\hspace{-.5cm}F = \int d^3R \Big[
\frac{7m^2 \vf^3\zeta(3)c_m}{96\pi^4 T^2} |\grad \Delta|^2 + V_2 |\Delta|^2 
\\
&& \qquad  \qquad \qquad  \qquad 
+ \frac{1}{2} 
V_4 |\Delta|^4 + \frac{1}{3} V_6 |\Delta|^6 
\Big],
\nonumber
\eea
where $c_m \simeq 0.0789$ and we defined the Fermi velocity  $\vf = 2\pi^2 \rho_0/m^2$.

To reduce $F$ to the form of Eq.~(\ref{eq:freeoriginal}), we 
 define  $\Delta(\bx) = \eta \psi(\bx)$
with $\eta \simeq 3.3\big(\frac{\pi^2 T^2}{\rho_0 \mu \zeta(3)}\big)^{1/2}$ chosen to fix the 
coefficient of the gradient term.  Then we'll have $r = 2V_2 \eta^2$ for the quadratic coefficient,
$u = 2V_4 \eta^4$ for the quartic coefficient, and $v = 2V_6 \eta^6$ for the sixth order coefficient.
Near the tricritical point at unitarity, these are approximately given by:
\bse
\bea
\label{eq:r}
r &\simeq & 2\rho_0 \eta^2 (t+ 0.86h) ,
\\
u &\simeq  &-0.16 \frac{\rho_0 \eta^4}{\Tt^2} h, 
\\
v &\simeq & .0064 \frac{\rho_0\eta^6}{\Tt^4},
\eea
\ese
where $h \equiv \frac{H-\Ht}{\Ht}$ and $\rho_0 = \frac{m^{3/2} \mu^{1/2}}{\sqrt{2}\pi^2\hbar^3}$ is the 
density of states at $\mu$.   Thus, as already mentioned, $r$ vanishes along a line in the $T$-$H$ plane,
(that is the continuous superfluid to nonsuperfluid transition occuring for $T>\Tt$) and $u$ vanishes
at $H\to \Ht$.  The sixth-order coefficient is approximately
constant near the TP. 

\section{Renormalization Group}
\label{SEC:RG}
Having provided an approximate connection between the microscopic one-channel model Eq.~(\ref{eq:singlechannelintro})
and the free-energy functional  Eq.~(\ref{eq:freeoriginal}), we now proceed to analyze the 
enhanced critical fluctuations near the tricritical point of $F$. 
Such fluctuations, and the concomitant logarithmic corrections to scaling~\cite{Larkin69,WR72,Stephen75,Lawrie}, are best analyzed 
using the renormalization group (RG), which incorporates the effect of fluctuations
neglected in the mean-field approximation.  It is convenient for such an analysis to 
set $\hbar^2/m_b = 1$ (equivalent to measuring lengths in different units), so that 
the gradient term of  Eq.~(\ref{eq:freeoriginal}) has coefficient $1/2$:

\be
F = \int d^3 x \big[
\frac{1}{2} |\grad \psi|^2 + \frac{1}{2} r  |\psi|^2  +  \frac{1}{4} u  |\psi|^4 + \frac{1}{6} v  |\psi|^6
\big].\label{eq:triham}
\ee

   We first note that the model 
Eq.~(\ref{eq:triham}) is defined for momenta below an upper cutoff $\Lambda$,
i.e., it is coarse-grained on length scales larger than $\sim \hbar/\Lambda$. (For a unitary 
polarized Fermi gas, we expect $\Lambda \sim \kf$, with $\kf$ the Fermi wavevector, on dimensional grounds.)
  The perturbative RG proceeds by integrating out states close to $\Lambda$.  Thus, 
we split the Fourier transform $\psi(\bp) = \int d^3 x\, {\rm e}^{i\bp\cdot \bx} \psi(\bx)$ into 
low and high momentum modes
 $\psi(\bp) = \psi_<(\bp) + \psi_>(\bp)$ with $\psi_<(\bp)$ defined for $0<p<\Lambda/b$  
and $\psi_>(\bp)$ 
defined for the shell of momenta $\Lambda/b<p<\Lambda$ with $b>1$.  By evaluating the trace over $\psi_>(x)$, perturbatively in
$u$ and $v$ (focusing on the crucial $\ln \Lambda$-divergent terms), 
we derive an effective theory for $\psi_<(x)$ that is of the form of the original model but with renormalized couplings.

Let's illustrate this procedure for the contribution to the effective Hamiltonian from expanding the sixth-order
term to $\curO(v^2)$.  This is:
\bea
&&F_{v^2} = -\frac{1}{2} \frac{v^2}{36} \int d^3 x_1  d^3 x_2 \langle 
|\psi_<(x_1) + \psi_>(x_1)|^6
 \\
&&\qquad \qquad \qquad \qquad \qquad \qquad 
\times |\psi_<(x_2) + \psi_>(x_2)|^6
\rangle_>,\nonumber
\eea
where the subscript $>$ indicates the trace over the high-momentum modes.  This trace is evaluated with the help of
the two-point Green function 
\be
\langle \psi_>^\dagger(x_1) \psi_>(x_2) \rangle = 2 G_>(x_1 - x_2),
\ee
which has the Fourier transform 
\be
G_>(p) = \frac{1}{p^2 + r},
\ee
for $\Lambda/b<p<\Lambda$.  This yields:
\be
 F_{v^2} = - 224 \frac{v^2}{6}  \int d^3 x |\psi_<(x)|^6 \int d^3 x' G_>(x')^3,
\label{eq:hvsquared}
\ee
so that the $\curO(v^2)$ term clearly renormalizes the bare sixth-order coupling $v$.  To proceed, we must evaluate
the final integral in Eq.~(\ref{eq:hvsquared}).  If the momenta appearing in $G(p)$ were unrestricted, then this integral 
would be $\ln \Lambda$ divergent, and given by:
\bea
I(\Lambda) &\equiv &\int\frac{d^3 p_1}{(2\pi)^3} \int\frac{d^3 p_2}{(2\pi)^3} \frac{1}{p_1^2+r}\frac{1}{p_2^2+r}\frac{1}{(\bp_1+\bp_2)^2+r},
\nonumber
\\ &\simeq & \frac{1}{16\pi^2}\ln \frac{\Lambda}{r}, 
\eea
where we took $r>0$ for simplicity.
In the present case, all momenta are restricted to the high-momentum window $\Lambda/b<p<\Lambda$; for this case we find
\be
\int d^3 x' G_>(x')^3 \simeq I(\Lambda) - I(\Lambda/b) =  \frac{1}{16\pi^2}\ln b,
\ee
thus yielding
\be
 F_{v^2} = - \frac{14}{\pi^2} \ln b \frac{v^2}{6} \int d^3 x |\psi_<(x)|^6.
\ee
To complete the RG procedure, we must define new momenta $p' = bp$ to restore the original cutoff 
$\Lambda$, equivalent to defining $x' = b^{-1}x$.  As usual, this also causes a rescaling of the kinetic
energy term in Eq.~(\ref{eq:triham}) that must be absorbed into a new fermion field $\psi(\bx')$:
\be
\psi_<(b\bx') = b^{-\frac{1}{2}} \psi(\bx').
\label{eq:wfrenorm}
\ee
With these manipulations, the final effective Hamiltonian is exactly of the form of Eq.~(\ref{eq:triham})
but with the renormalized coupling 
\be
v'= v- c v^2 \ln b,
\ee
with the numerical coefficient $c =\frac{14}{\pi^2}$.  Upon iterating the RG procedure, and including
similar  renormalizations for the quadratic and quartic coefficients, we find the RG equations:
\bse
\label{eq:allrg}
\bea 
\frac{dv(b)}{d\ln b} &=& -c v(b)^2,
\label{eq:vrg}
\\
\frac{du(b)}{d\ln b} &=& u(b)[1 -  c_4 v(b)],
\label{eq:urg}
\\
\frac{dr(b)}{d\ln b} &=& 2r(b)- c_2u(b)^2,
\label{eq:rrg}
\eea
\ese
for the running coupling constants $u(b)$, $v(b)$ and $r(b)$,
with   $c_4 = \frac{6}{\pi^2}$ and $c_2 = \frac{1}{2\pi^2}$.  These RG equations, consistent with previous results
reported in Refs.~\onlinecite{Stephen75,Lawrie},
can be integrated to yield
\bse
\label{eq:RGs}
\bea
v(b) &=& v\big(1+c v\ln b\big)^{-1},
\\
u(b) &=& ub\big(1+c v\ln b\big)^{-3/7},
\label{eq:uresult}
\\
r(b) &=& b^2\big(
r+\frac{1}{4}\frac{u^2}{v}\big[1-(1+cv\ln b)^{\frac{1}{7}}]
\big), \label{Eq:rofb}
\eea
\ese
with the latter equation being given by  $r(b) \simeq b^2 r$ 
close to the tricritical point.

\section{Molecular density}
\label{SEC:molecular}
Having computed the RG equations, our next task is to combine these with
scaling relations for various experimentally-observable quantities. 
We begin with the molecular density~\cite{densitynote} $n_m = |\psi|^2$, which satisfies the scaling 
equation
\be
n_m(r,u,v) = b^{-1} n_{m}(r(b),u(b),v(b)),
\label{eq:renormden}
\ee
following from Eq.~(\ref{eq:wfrenorm}). Below, we use the shorthand $n_m =b^{-1}n_{m{\rm R}}$ for such an equation.
 The left side of Eq.~(\ref{eq:renormden}) is the physical molecular 
density, while the right side is the density in the renormalized system. 
 The RG strategy is quite simple:
Although mean-field theory is invalid close to criticality, we can choose $b =b_*$ such that the {\it renormalized}
system is far from criticality where mean-field theory is accurate. 
 This is seen directly from Eq.~(\ref{eq:RGs}):
with increasing $b$, $|u(b)|$ and $|r(b)|$ grow large while $v(b) \to 0$ (validating perturbation theory). 
  In the critical and tricritical regimes, it
is sufficent to take the condition $|r(b_*)| \simeq 1$ which yields $b_* \simeq 1/|r|^{1/2}$ for the RG 
condition.  Then, using the 
mean-field result [i.e., Eq.~(\ref{eq:themeanfieldresult})] for the right side of Eq.~(\ref{eq:renormden}),
we find for the molecular density:
\be
n_m \simeq \frac{u\ell_r^{\frac{4}{7}}}{2v} \Big[\sqrt{1+ \frac{4|r|v}{u^2}\ell_r^{-\frac{1}{7}}}-1\Big],
\label{eq:nmresult}
\ee
where 
\be
\ell_r \equiv 1+ c v \ln |r|^{-\frac{1}{2}},
\ee
is a logarithmic correction to scaling.  
Equation~(\ref{eq:nmresult}) describes the molecular density near the tricritical point along
lines that intersect the critical line or the tricritical point. 
In the 
tricritical regime $H\to \Ht$, where $u\to 0$, this yields Eq.~(\ref{eq:nmintro}) for the leading 
temperature dependence.  As noted in Sec.~\ref{SEC:model}, Eq.~(\ref{eq:nmresult})
implies a distinctly different spatial dependence of the molecular density in a harmonic trap, seen
in Fig.~\ref{fig:densityplot}, with a spatial variation described by Eq.~(\ref{Eq:cloudnearedge}).

    For any $T>\Tt$, above from the tricritical point, 
Eq.~(\ref{eq:nmresult}) crosses over asymptotically to critical behavior 
\be
n_m \simeq \frac{|r|}{u} \ell_r^{3/7},
\label{Eq:densitycritical} 
\ee
therefore yielding a standard TF profile, with $n_m \propto (1-x^2/x_{TF}^2)$ near the edge.
However, close to the TP, this crossover will only occur extremely close to the edge of the cloud.  As the 
tricritical point is approached from the critical regime $u>0$, the slope of the boundary is predicted
to diverge as $u\to 0$ according to Eq.~(\ref{Eq:densitycritical}), as seen in Fig.~\ref{fig:densityplot}.
 This vanishing of $u$ can be interpreted
as a vanishing~\cite{SR2006} of the {\it effective\/} molecule-molecule scattering length  $a_m$ (recently 
measured~\cite{ShinPreprint}); using  Eq.~(\ref{eq:uresult}) along with the above renormalization 
condition gives,  at $T=\Tt$ and for $h\to 0$, 
\be
a_m \propto |h|\ln^{-3/7}(|h|^{-1}),
\ee
for the vanishing of the molecular scattering length close to the TP.

We now turn to the behavior of the molecular density across the phase transition in the first-order regime
($T<\Tt$) near the TP.  As noted above, in the continuous regime near the TP, the superfluid order parameter exhibits
a sharp onset at the phase transition, resulting in the steep edge of the cloud shape shown in Fig~\ref{fig:densityplot}.
In the first-order regime, the order parameter jumps discontinuously across the transition, resulting in a jump
in the molecular density at the edge of the molecular cloud.

The molecular density  in the first-order regime can similarly be computed using the RG; in this case, 
however, we must choose a more generally-valid renormalization condition~\cite{NelsonRudnick}.  
We'll take this to be  when the curvature of the free-energy minimum  
$\frac{d^2F}{d|\psi|^2}$ in the ordered state is large (i.e., order unity); this yields
\be
1 = \frac{1}{v_*}\big[ u_*^2 - 4r_*v_* - u_*\sqrt{u_*^2-4r_*v_*}\big],
\label{eq:generalcondition}
\ee  
for the most general condition in the ordered state, with $v_* \equiv v(b_*)$, etc.  Close to the 
continuous transition or to the tricritical point, Eq.~(\ref{eq:generalcondition}) for $b_*$ is essentially equivalent to 
$b_* \simeq 1/|r|^{1/2}$ as used above.
 In the first-order regime $u<0$,  however, Eq.~(\ref{eq:generalcondition}) is, 
approximately, 
\be
1 \simeq 2u_*^2/v_*,
\label{eq:generalconditionapprox}
\ee
giving our renormalization condition in the first-order regime. 
   To compute the jump in the order parameter or molecular density across
the phase boundary, we need use this condition to find the first-order phase boundary, which in mean-field theory
is given by $r = \frac{3}{16} \frac{u^2}{v}$.  Combining this with Eq.~(\ref{eq:generalconditionapprox})
then yields 
\be
r = \frac{3u^2}{16v}\ell_{2u^2/v}^{\frac{1}{7}}
\label{Eq:renormphasebound}
\ee
for the {\it renormalized\/} first-order phase boundary. 
Along  with Eq.~(\ref{eq:renormden}), we can then obtain the jump 
\be
\delta n_m  \simeq \frac{3|u|}{4v} \big( 1+ c v \ln \frac{\sqrt{v}}{\sqrt{2} |u|}\big)^{\frac{4}{7}},
\label{eq:densityjump}
\ee
in the molecular density across the first-order phase boundary, which vanishes as $u\to 0$ at the TP as expected.

\section{Magnetization}
\label{SEC:mag}

We now turn to the local polarization, or magnetization, $M = n_\uparrow - n_\downarrow$,
a crucial observable in present-day experiments, that is related to the free energy via $M = -\frac{\partial F}{\partial H}$.
In deriving Eq.~(\ref{eq:freeoriginal}), we dropped the overall constant term $F_0$ in the free energy;
we must now reinstate this term which yields a fermion-only contribution to the magnetization that we denote $M_f$.  Thus, we have
\be
M = M_f - \frac{1}{2} \frac{\partial r}{\partial H}|\psi|^2 -
 \frac{1}{4} \frac{\partial u}{\partial H}|\psi|^4 -
 \frac{1}{6} \frac{\partial v}{\partial H}|\psi|^6  ,
\label{eq:m}
\ee
for the magnetization.  

Near the tricritical point where $|\psi|$ vanishes, $M_f$ will be the largest of the terms in
Eq.~(\ref{eq:m}) but not show any significant signature of the phase transition. However, one observable that will
show a sharp signature across the phase transition is the jump in the magnetization across the phase boundary
in the first-order regime.  This quantity, that is directly measurable as the jump in magnetization as 
a function of radius in a trapped polarized Fermi gas~\cite{Shin07}, is equivalent to the width of the coexistence
region $\delta M \equiv M_{c2}-M_{c1}$ below the tricritical point, see Fig.~\ref{phasediagram}b.

The $\curO(|\psi|^2)$ term provides the dominant contribution to  $\delta M$.  Using Eq.~(\ref{eq:densityjump})
for the jump in $|\psi|^2$ across the phase boundary along with Eq.~(\ref{Eq:renormphasebound}), we obtain 
\be
\delta M \simeq \frac{\sqrt{3}}{2} \frac{\partial r}{\partial H}\sqrt{\frac{r}{v}} \big( 1+ c v \ln r^{-1/2})^{1/2},
\ee
for the jump in the magnetization in the first-order regime near the TP.

\section{Heat Capacity}
\label{SEC:heat}
In the present section, we use the RG to obtain corrections to mean-field theory for the free energy
and heat capacity $C = -T\frac{d^2F}{dT^2}$.  Close to the phase transition, we can take 
$C\propto \frac{\partial^2 F}{\partial r^2}$, since the leading $T$ dependence is via $r$.
Using the free-energy scaling relation $F = b^{-3} F_{\rm R}$, we find 
\be
C\big(r,u,v\big)= bC\big(r(b),u(b),v(b)\big),
\label{eq:heatrg}
\ee
 for the RG equation
for the heat capacity.    We begin with the heat capacity in
the ordered phase.  In this regime, we'll use the mean-field result for the heat capacity for
the right side of Eq.~(\ref{eq:heatrg}), which is obtained by differentiating the mean-field
free energy Eq.~(\ref{eq:meanfieldfree}) and using the stationarity condition 
Eq.~(\ref{eq:stationarity}):
\be
C \simeq - \Tt\big(\frac{\partial r}{\partial T} \big)^2\frac{1}{2}\frac{d|\psi|^2}{dr} = T \big(\frac{\partial r}{\partial T} \big)^2\frac{1}{2}\frac{1}{\sqrt{u^2+4|r|v}}.
\ee
Taking the RG condition  $b_*  = 1/\sqrt{|r|}$ (appropriate for the critical and tricritical regimes, on which 
we shall focus), we finally obtain in the {\it ordered\/} state $r<0$:
\be
C \simeq \Tt \frac{1}{2}\big(\frac{\partial r}{\partial T} \big)^2 \frac{1}{\sqrt{u^2\ell_r^{-\frac{6}{7}} + 4v |r|\ell_r^{-1}}},
\label{eq:generalheat}
\ee 
for the fluctuation contribution to the heat capacity
which, for $H =\Ht$ on a line crossing the TP, gives
\be
C\simeq  \Tt \frac{\sqrt{c}}{4}\big(\frac{\partial r}{\partial T} \big)^2 \frac{\sqrt{\ln |r|^{-1/2}}}{\sqrt{|r|}},
\label{Eq:cordered}
\ee
that reduces to the previously quoted formula for the tricritical heat capacity, Eq.~(\ref{eq:heatintro}),
once we use $r\propto t$ [as follows from Eq.~(\ref{eq:r})]

In the critical region above the TP (for $T>\Tt$), Eq.~(\ref{eq:generalheat}) gives~\cite{criticalnote} 
$C\propto (2u)^{-1}\ell^{3/7}_r$.  In the absence of the log correction, this would represent the usual mean-field specific 
heat jump at a superfluid transition; however, the true asymptotic behavior  in this regime is known to reflect a very small heat capacity
exponent $\alpha$~\cite{Barmatz}.

In the disordered (normal) state, the mean-field free energy vanishes (since $\psi = 0$ there). Thus, to obtain a nonzero result
for $C$ at $r>0$ we must go beyond mean-field theory.  The free-energy in the normal state, $\curF$,  is given by a functional
integral over the field $\psi$ 
\be
\curF = - T \ln \int D\psi\, {\rm e}^{-F/T},
\ee
with the action $F$ given in Eq.~(\ref{eq:triham}).  Near the transition, we can assume that the
leading $T$ dependence comes through $r$, which yields
\be
C\simeq - \Tt\big(\frac{\partial r}{\partial T} \big)^2\frac{1}{2}\frac{d}{dr} \langle |\psi|^2  \rangle,
\ee
with the angle brackets reflecting the thermodynamic average with respect to $F$.  We can obtain 
the leading-order approximation to 
 $\langle |\psi|^2  \rangle$ using the normal-state Green function:
\be
\langle |\psi|^2  \rangle = \int \frac{d^3k}{(2\pi)^3} \frac{2}{k^2+r},
\ee
which, upon differentiating, gives for the heat capacity in the normal state:
\be
C \simeq \Tt 
\big(\frac{\partial r}{\partial T} \big)^2 \frac{1}{8\pi \sqrt{r}}.
\label{eq:cnormaldis} 
\ee
Using this result for the right side of Eq.~(\ref{eq:heatrg}), and taking $r(b_*)$ to be given by Eq.~(\ref{Eq:rofb}), 
we obtain for the heat capacity in the normal state ($r>0$):
\be
C \simeq \Tt 
\big(\frac{\partial r}{\partial T} \big)^2 \frac{1}{8\pi} \frac{1}{\sqrt{r + \frac{1}{4}\frac{u^2}{v} (1-\ell_r^{1/7}) }},
\ee
which, along a line intersecting the tricritical point ($u=0$), immediately reverts to Eq.~(\ref{eq:cnormaldis}), 
i.e., we find no logarithmic corrections in the normal state at the tricritical point so that, for $H=\Ht$,
 \be
C \propto \frac{1}{\sqrt{t}},
\ee
in the normal (nonsuperfluid) phase. 

\section{Correlation length}
\label{SEC:corr}
Finally, we consider the divergence of the correlation length $\xi$ near the phase transition, governed by
the decay of the superfluid correlation function at large distances.  Here we shall focus only
on a line crossing the tricritical point, i.e., we take $u = 0$.  In the normal phase, we have
\be
\langle \psi^\dagger(x)  \psi(0) \rangle  = 2G(x),
\ee 
with the Green function satisfying the scaling relation $G(x) = b^{-1} G_R(b^{-1}x)$ 
[as follows from Eq.~(\ref{eq:wfrenorm})]. Using the perturbative result for $G_R(x)$ and the
renormalization condition $b_* = 1/\sqrt{r}$ gives 
\be
\langle \psi^\dagger(x)  \psi(0) \rangle   = \frac{1}{2\pi x}{\rm e}^{-\sqrt{r}|x|},
\ee
identical to the perturbative result 
[due to the trivial scaling of $r$ near the TP Eq.~(\ref{Eq:rofb})], which, upon examining the argument of
the exponential gives the correlation length $\xi \propto r^{-1/2}$, and the exponent $\nu = \frac{1}{2}$.
A similar analysis in the ordered phase (expanding the free-energy around the mean-field solution) gives
$\xi \propto |r|^{-1/2}$ below the TP.  

\section{Concluding remarks}
\label{SEC:concl}
To conclude, we have computed various experimental predictions
for the behavior of a polarized superfluid Fermi gas near its tricritical point using  a GL
functional that is generally valid near the tricritical point even in
the unitary regime.  We presented results for numerous observables in cold-atom experiments, including
the onset of the superfluid order parameter at the transition, the magnetization jump across the first-order phase boundary, the heat capacity, 
and the correlation length.
In contrast to the standard superfluid transition, in which critical exponents deviate from their mean-field values
(exhibiting anomalous values), for tricritical points the mean-field exponents are predicted to be exact, but 
with logarithmic corrections.  
These predictions should provide sharp signatures of the tricritical point in polarized Fermi gases.

%----------
\smallskip
\noindent
{\it Acknowledgments\/} ---  
%---------
We gratefully acknowledge discussions with  W. Ketterle, A. Lamacraft, L. Radzihovsky, and  Y. Shin,  and 
the Aspen Center for Physics where part of this work was carried out.
This research was supported by the Louisiana Board of Regents,
under grant No. LEQSF (2008-11)-RD-A-10. 
%--------------------------


\begin{thebibliography}{10}
\bibitem{densitynote}
We note here that our identification of the square of the order parameter $|\psi|^2$ with the molecular density
is strictly only valid in the deep BEC regime, and will break down as the unitary point is approached.
\bibitem{criticalnote}
The logarithmic corrections obtained here are strictly only valid at the
tricritical point; our results in the second-order regime must eventually cross over
to 3D XY behavior.
\end{thebibliography}
\end{document}